\documentclass[12pt,preprint]{aastex}
\everymath{\rm}
\everydisplay{\sf}
\received{}
\revised{}
\accepted{}
\cpright{}{}\journalid{}{}
\articleid{}{}
\paperid{}
\slugcomment{To appear in {\it The Astrophysical Journal Supplement}}
\shortauthors{Milingo, Kwitter \& Henry}
\shorttitle{S, Cl, \& Ar in PNe}
\begin{document}
\title{Sulfur, Chlorine, and Argon Abundances in Planetary Nebulae. IIA: Observations of a Southern Sample}

\author{J.B. Milingo\footnote{Visiting Astronomer, Cerro Tololo Interamerican Observatory, National Optical Astronomy Observatories,
which is operated by the Association of Universities for Research in
Astronomy, Inc. (AURA) under cooperative agreement with the
National Science Foundation.}}

\affil{Department of Physics, Gettysburg College, Gettysburg, PA  17325; jmilingo@gettysburg.edu}  

\author{K.B. Kwitter}

\affil{Department of Astronomy, Williams College, Williamstown, MA
01267; kkwitter@williams.edu}

\author {R.B.C. Henry$^1$}

\affil{Department of Physics \& Astronomy, University of Oklahoma,
Norman, OK 73019; henry@mail.nhn.ou.edu}

\and

\author {R.E. Cohen\footnote{Keck Northeast Astronomy Consortium Summer Fellow at Williams College, 2001}}

\affil{Department of Astronomy, Wesleyan University, Middletown, CT 06459; recohen@wesleyan.edu}

\begin{abstract}

In this paper we present fully reduced and dereddened emission line
strengths for a sample of 45 southern Type~II planetary nebulae (PNe).
The spectrophotometry for these PNe covers an extended optical/near-IR
range from 3600 - 9600~\AA. This PN study and subsequent analysis
(presented in a companion paper), together with a similar treatment
for a northern PN sample, is aimed at addressing the lack of
homogeneous, consistently observed, reduced, and analyzed data sets
that include the near-IR [S~III] lines at 9069 and 9532 \AA. The use
of Type~II objects only is intended to select disk nebulae
that are uncontaminated by nucleosynthetic products of the progenitor
star. Extending spectra redward to include the strong [S~III]
lines enables us to look for consistency between S$^{+2}$ abundances
inferred from these lines and from the more accessible, albeit weaker,
[S~III] line at $\lambda$6312.

\end{abstract}


\clearpage

\section{Introduction}

The study of the origin and buildup of the chemical elements in the universe
relies heavily on the availability of accurately measured abundances for
constraining model predictions. This paper is the second in a series that
endeavors to measure abundances of S, Cl, and Ar, all relative to O, and to
compare our empirical results with model calculations available in the
literature.

One of the salient features of this project is to provide a large
set of carefully and consistently measured abundances based upon a
homogeneous, spectrophotometric data set which extends from 3600-9600 {\AA}.
An important asset of this extended coverage is that it includes the strong
nebular lines of [S~III] $\lambda\lambda$9069,9532 which have never before
been used extensively to study the S abundances in planetary nebulae (PNe).
Previously, S$^{+2}$ abundances have been inferred primarily by using the
auroral line of [S~III] at $\lambda$6312, but this line is naturally weak and
extremely temperature-sensitive, raising some doubts about the accuracy of the
ion abundances derived. The inclusion of both the auroral and nebular lines of
[S~III] allows us to compute S$^{+2}$ abundances from both line types and to
comment on the common use of the auroral line by other investigators to
study sulfur abundances.

Kwitter \& Henry (2001; hereafter Paper~I) observed 19 northern Type~II PNe
over the same spectral region as described above. The reader is referred to their
paper for a more extensive discussion of the motivation for this project. The
current paper, Paper~IIA in the series, presents spectrophotometric
measurements for 45 southern Type~II PNe; the analysis of the data and
calculation of abundances based upon the data will follow in Paper~IIB
(Milingo, Henry, \& Kwitter, in preparation). The current study, then, is
complementary to Paper~I. Future papers in the series will consider
objects for which spectroscopic measurements are available in the IR from the
ISO satellite along with a final interpretation of abundance patterns observed
within the entire sample.

In the next section, we present the details of the observations and data
reductions ({\S}2.1), along with a careful consideration of the accuracy of the
measured line strengths ({\S}2.2). A brief summary is given in {\S}3.

\section{Data}

\subsection{Observations}

Observations were obtained at CTIO in 1997 from 29 March - 3 April
using the 1.5m telescope and cassegrain spectrograph with Loral 1K
CCD.  The 1200 x 800 Loral 1K CCD has 15$\mu$ pixels.  We used a
5$\arcsec$ x 320$\arcsec$ extended slit in the E-W direction which
accommodated most of the the angularly small objects in this
collection.  Perpendicular to dispersion the scale was
1.3$\arcsec$/pixel.  Gratings \#22 and \#9 were used to obtain
extended spectral coverage from 3600-9600 \AA  with overlap in the
H$\alpha$ region.  Both gratings have nominal wavelength dispersions
of 2.8 \AA/pixel and 8.6 {\AA}~FWHM resolution.  Table 1 lists the
objects observed, their angular sizes in arcseconds, and the exposure
times in seconds for the blue and red spectral regions observed with
gratings \#9 and \#22 respectively.  Most of the PNe were observed
centered along the slit.  For the angularly small objects, the central
star was unavoidably included.  In the case of a few angularly large
objects we centered on the brightest portion of the nebula.  Nightly
calibration frames were obtained including bias exposures, twilight
flats, dome flats, HeAr comparison spectra for wavelength calibration
and standard star spectra for flux sensitivity calibration.
 
The Loral 1K CCD produces significant fringing at wavelengths beyond
$\sim$7500 \AA.  Assuming less than ideal fringe removal via dome
flats, reported amplitudes for interference fringes are $\pm$1.25\% at
7500 \AA, $\pm$2.25\% at 8000 \AA, $\pm$5.7\% at 8500 \AA, $\pm$8\% at
9000 \AA, and at the longest wavelengths we measure $\pm$9.8\% at 9500
\AA.  We note this contribution to the uncertainty in our line
intensities measured at wavelengths longer than 7500 \AA.

The original two-dimensional spectra were reduced and calibrated using
standard long-slit spectrum reduction methods in IRAF\footnote{IRAF is
distributed by the National Optical Astronomy Observatories, which is
operated by the Association of Universities for Research in Astronomy,
Inc. (AURA) under cooperative agreement with the National Science
Foundation.}.  One-dimensional spectra were extracted from the
original two-dimensional images interactively using the {\it kpnoslit}
package. Line fluxes were measured with {\it splot}. The final
calibrated, merged spectrum of IC~2621 is shown in Fig. 1 as an illustration
of our data; important emission lines are identified.

\subsection{Line Strengths}

The focus of this paper is the listing of line strengths presented in
Tables~2A-H. Each table shows the line identification and wavelength
in the first column. Fluxes uncorrected for reddening are presented in
columns labeled F($\lambda$), where these flux values have been
normalized to H$\beta$=100 using our observed value of log
F$_{H\beta}$ shown in the last row of the table.  These line strengths
in turn were corrected for reddening by assuming that the dereddened
relative strength of H$\alpha$/H$\beta$=2.86 (Hummer \& Storey 1987,
N$_e$=100 cm$^{-3}$, T$_e$=10,000 K) and computing the logarithmic
extinction quantity $c$ shown in the penultimate line of the table.
Values for the reddening coefficients, f($\lambda$), are listed in
column~(2), where we employed the extinction curve of Savage \& Mathis
(1979). Intensities, I($\lambda$), were obtained by multiplying
F($\lambda$), the observed ratio relative to H$\beta$, by
dexp[$c$f($\lambda$)], where dexp[y]=10$^y$. In general, intensities
of strong lines, i.e. those whose strengths are equal to or greater
than that of H$\beta$, have measurement uncertainties $\le$10\%;
single colons indicate uncertainties of more than $\sim$25\%, and
double colons denote uncertainties exceeding $\sim$50\%. This includes
estimated errors in the $c$ values of $\sim$0.15. Our uncertainty
estimates are based on extensive previous experience with similar
types of measurements. However, we also considered consistency between
measurements performed independently on the data by two of us (JBM and
KBK), as well as the quality of agreement between theory and
observation for line ratios which have theoretically determined
values. (See discussion of Table~3B below.)

As a check on our reddening correction method, we have also calculated the
extinction quantity, $c$, based on both the Paschen-10 and Paschen-8 lines at
$\lambda$9014 and $\lambda$9546, respectively. Results of this exercise are
presented in Table~3A, where for each object given in the first column we list
the values of $c$(H$\alpha$), $c$(P10), and $c$(P8). All three values were
calculated based upon theoretical line ratios given in Pengelly (1964).
Since $c$ is a measure of extinction column length, in principle all three
values of $c$ for any one object should be the same.

Figure~2 is a plot of c(P10) (open circles) or c(P8) (open squares)
versus $c$(H$\alpha$) for objects for which these values could be
computed. The solid diagonal line indicates the track for the ideal
one-to-one ratio. With the exception of a few outliers, agreement is
satisfactory, although there is a suggestion of a systematic trend in
which $c$ for the Paschen lines tends to be slightly lower for a given
c(H$\alpha$).  This effect could arise if telluric absorption in the
region of the Paschen lines suppresses these lines, causing the values
of $c$ derived from them to be smaller. The same effect would result
if H$\alpha$ were systematically overestimated due to [N~II]-H$\alpha$
deblending problems. It should also be noted that the deconvolution of
P8 from [S~III] $\lambda$9532 was difficult for most of these objects
for two reasons. First, the ratio of P8/$\lambda$9532 is in many cases
only a few percent. Second, this blend falls near the edge of the
spectrum where the focus is poorer, leading to broadened lines which
are harder to deconvolve.

An additional check for data quality is provided by information in
Table~3B, where for each object we list the measured values of seven
line ratios that are set solely by atomic parameters and thus are
unaffected by environmental conditions. The ratios are defined in the
table footnote. Values for each ratio were computed from dereddened
line strengths in Tables~2. We omitted ratios where either one of the
lines is marked with a double colon in Table~2. For [Ne~III], our
spectral resolution was insufficient to explicitly separate [Ne~III]
$\lambda$3968 from H$\epsilon$; we therefore calculated the
H$\epsilon$ intensity from the theoretical ratio of
H$\epsilon$/H$\beta$ in Hummer \& Storey (1987), and subtracted it from the
blend. Note that the collection of intrinsically-constant ratios in
Table~3B samples regions all across our spectral coverage. The
penultimate line in the table gives the average for the column above
it, while the last line shows the theoretical value. In a perfect
world the theoretical value and all the numbers above it would agree
precisely; the level of imperfection is indicated by the standard
deviations associated with the means and by a general inspection of
all of the numbers in any one column. There is good agreement for all
line ratios.

Finally, while we do not explicitly list them here, several objects
appearing in the present sample were also a part of the northern
sample discussed in Paper~I. A comparison of the raw line strengths
and derived $c$(H$\alpha$) values between these two samples shows
very good agreement. Given that the two sets of data were
observed, reduced, and measured completely independently by different
individuals using different instruments, this favorable agreement
increases our confidence in both our
method and in the resulting line strengths in each sample.

\section{Summary}

We have described spectrophotometric observations between 3600-9600 {\AA} and
presented an extensive list of line strengths for a sample of 45 southern
Type~II planetary nebulae. In a companion work, Paper~IIB (Milingo, Henry, \&
Kwitter 2002) we will use these data to derive electron temperatures and
densities, as well as ionic and elemental abundances. These two papers
are part of a large study of PNe described in detail in Paper~I, whose purpose
is to produce accurate measurements of interstellar S/O, Cl/O, and Ar/O ratios
and to compare these with predictions from theoretical models of stellar
nucleosynthesis.

\acknowledgments

We are grateful to the CTIO TAC for granting us observing time and to
the IRAF staff for their ready answers. Our research is supported by
NSF grant AST-9819123.

\clearpage



\clearpage

\begin{figure}
\figurenum{1}
\plotone{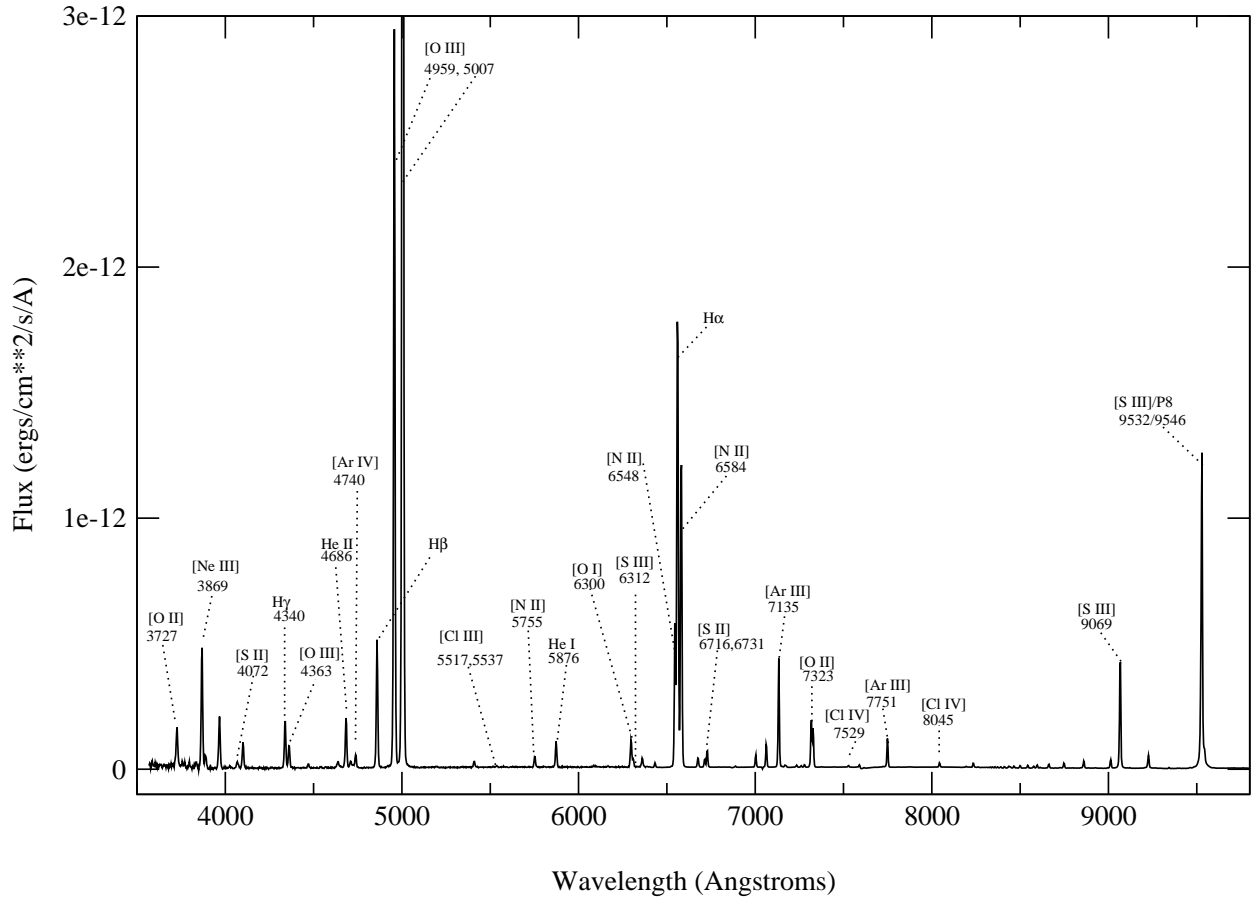}
\caption{Spectrum of IC 2621.}
\end{figure}

\begin{figure}
\figurenum{2}
\plotone{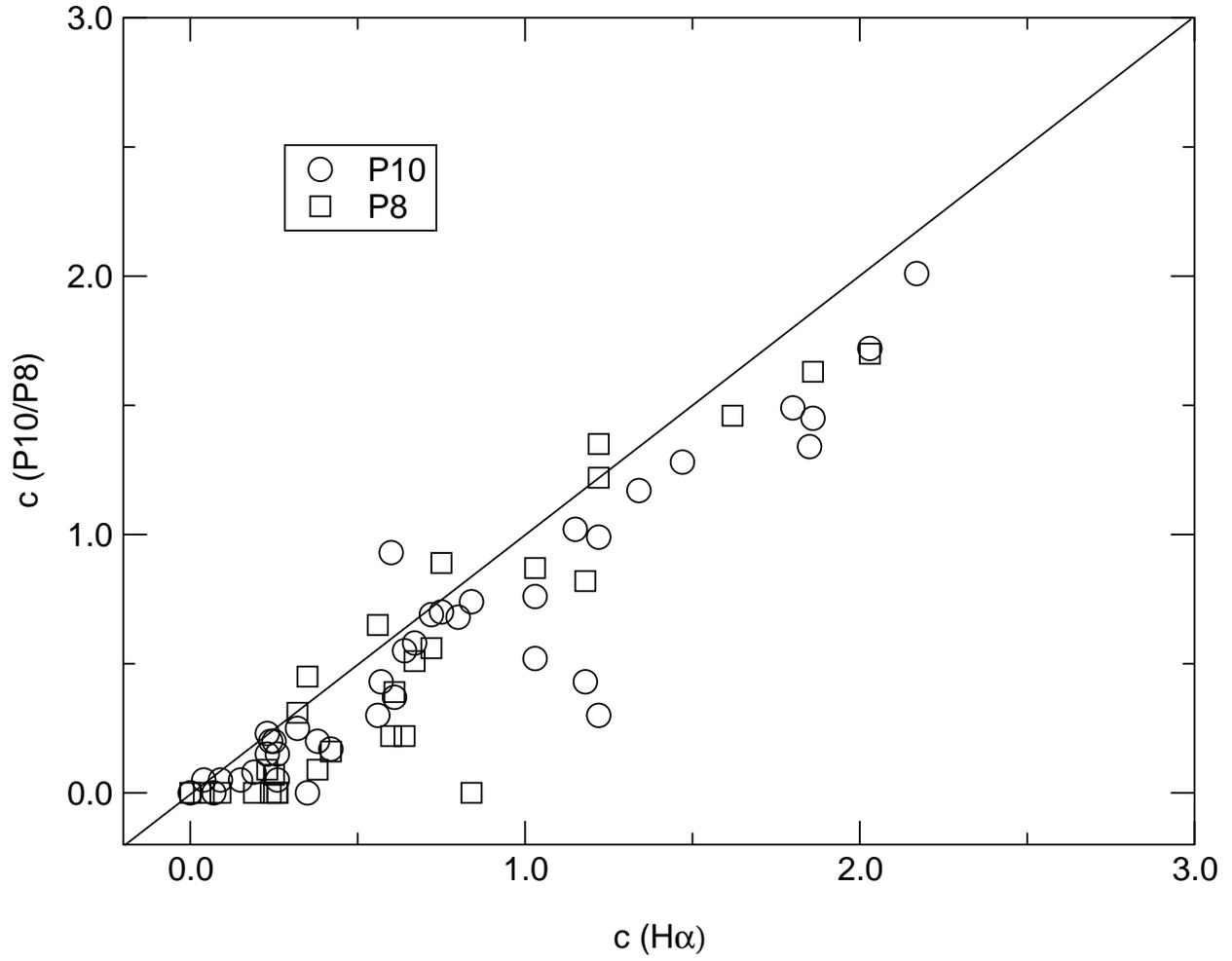}
\caption{A comparison of logarithmic extinction {\it c} as determined
using the Paschen 8 and 10 lines versus the value inferred from using
H$\alpha$. The solid line shows the track for a one-to-one
correspondence.}
\end{figure}

\end{document}